\documentclass[twocolumn,showkeys,showpacs,preprintnumbers,amsmath,amssymb]{revtex4}
\usepackage{graphicx}

\def\dd{\displaystyle}

\begin{document}
\title{\bf Casimir Energy For a Massive Dirac Field in One Spatial Dimension: A Direct Approach}
\author{R. Saghian}
\author{M.A. Valuyan}
\thanks{Present Address: Semnan Branch, Islamic Azad University, Semnan, Iran\\ \texttt{Electronic Address: m.valuyan@semnaniau.ac.ir; m-valuyan@sbu.ac.ir}}
\author{A. Seyedzahedi}
\author{S.S. Gousheh}
\affiliation{%
Department of Physics, Shahid Beheshti University G.C., Evin, Tehran
19839, Iran}

\date{\today}
\begin{abstract}
In this paper we calculate the Casimir energy for a massive
fermionic field confined between two points in one spatial
dimension, with the MIT Bag Model boundary condition. We compute the Casimir energy directly by summing over the allowed modes. The method that we use
is based on the Boyer's method, and there will be no need to resort
to any analytic continuation techniques. We explicitly show the graph of the Casimir energy as a function of the distance between the points and the mass of the fermionic field.
We also present a rigorous derivation of the MIT Bag Model boundary
condition.
\end{abstract}

\maketitle

\section{Introduction}
More than $60$ years have passed from the time when H. B. G. Casimir
published his famous paper\,\cite{h.b.g.} on what has been called
the Casimir effect ever since. It remained relatively unknown for
over two decades, but from the early $70$s this effect has attracted
much attention. The Casimir energy is a pure quantum effect with
macroscopic manifestations. Generally, the Casimir energy is defined
as the difference between the vacuum energies in the presence and
the absence of any external boundary conditions or background
fields. Both of these energies are in general infinite. However the
difference between the two has almost invariably been calculated to
be finite. The Casimir effects have been calculated for a variety of
fields, geometries, number of spatial dimensions, and boundary
conditions (for a review see
\,\cite{burdag.book.,milton.book.,milton.paper.,Mostepanenko.}).
\par
Recently the Casimir effect has been studied in connection with many physical phenomena. For example this effect has been studied in the context of the phenomenological chiral bag models of the nucleon\,\cite{chiral.casimir.,linas.}. In such models
"the bag constant", $B$, is an input parameter to the theory. As is
well known, this constant is added to the Lagrangian
density in order to balance the outward pressure of the quarks
by the inward vacuum pressure $B$ on the surface of the bag\,\cite{milton.book.}. This constant can be related to the Casimir energy\,\cite{bag1.}. The Casimir effect for the String and the Superstring leads to string theories having critical dimensions. The string is a finite two-dimensional system with an infinite phonon spectrum, the sum of the zero-point fluctuations of which leads to exactly the same calculation as in the original Casimir effect\,\cite{LBrink.}. Moreover, the presence of the Casimir effects in many different phenomena in condensed matter and laser physics have been established both theoretically and experimentally\,\cite{condensed.,Krech.}. For example, the Casimir force is important in the development of microtechnologies that routinely allow control of separation between bodies smaller than $1\mu m$\,\cite{nano.application.,Rodrigues.,Lambrecht.}. An interesting application of the field theoretical models
with compact dimensions recently appeared in nanophysics. The long-wavelength description
of the electronic states in graphene can be formulated in terms of a Dirac-like theory in three dimensional space-time with the Fermi velocity playing the role of speed of light\,\cite{5.,Vincenzo.,Gonzalez.,Lee.,Sharapov.,Castro.}. Single-walled carbon nanotubes are generated by rolling up a graphene sheet to form a
cylinder and the background space-time for the corresponding Dirac-like theory has topology $R^{2}\times S^{1}$\,\cite{graphene.ref.}.
\par
In order to compute the Casimir energy, physically relevant boundary
conditions must be imposed. For example the Casimir energy for an
electromagnetic field is usually calculated with the boundary
conditions appropriate for conducting boundaries as in
\,\cite{burdag.book.,milton.book.,milton.paper.,Mostepanenko.,EM.,Valuyan.,Hacgan.,Jordan.}.
The Casimir energy for scalar fields has been investigated with
the Dirichlet boundary condition
\,\cite{burdag.book.,milton.book.,milton.paper.,Mostepanenko.,valuyanjadval.,Moazzemi.,Gousheh.,valuyan2.,Mhammadi.},
the Neumann boundary
condition\,\cite{milton.book.,Mostepanenko.}, the mixed
boundary condition\,\cite{mixed.}, and the Robin boundary
condition\,\cite{Robins.,Albuquerque.,someaspects.}.
\par
From this point on we concentrate on the Dirac field. Since the Dirac field
obeys a first order differential equation of motion, it is
impossible to use the aforementioned boundary conditions. Moreover, boundary conditions are in general more
disruptive to Dirac fields than boson fields because the equations of motion are first
order\,\cite{Jaffe.}. A proper
boundary condition for the Dirac field is the MIT Bag Model boundary
condition. It is usually said to imply that there is no flux of
fermions through the boundary, i.e. if $j^{\mu}$ denotes the current
of the Dirac field and $n^{\mu}$ is the normal unit vector to the
boundary, then $n_{\mu}j^{\mu}=0$. However, it implies an even stronger condition which is the absolute confinement of the
fermionic field. One could confine the fermionic wave function by an infinite
scalar potential. This model was considered by
Bogolioubov \,\cite{Bogolioubov.} and later developed as the MIT Bag
Model by A. Chodos, et al \,\cite{Chodos1.,Chodos2.} for hadrons. It
is important to mention that in general the computation of the
Casimir energy for a massive Dirac field turns out to be much more
difficult than the massless case. There has been relatively few
studies for the Casimir energy inside closed surfaces, and we
mentioned a few of them here. The Casimir energy has been calculated
for a spherical geometry for a massive\,\cite{Elizalde.} and
massless fermionic fields \,\cite{milton.book.,milton.paper.,hofmann.}. It
has also been recently calculated for a massless fermionic field
subject to the MIT Bag Model boundary condition confined inside a
three dimensional rectangular geometry \,\cite{maghale.}.
\par
As usual there are many more studies on the two parallel plates
geometry, and we shall concentrate on this problem from this point
on. The first computation of the Casimir energy for the Dirac field
was done by Johnson in $1975$ \,\cite{Massless.}. He computed this
energy per unit area for a massless fermionic field subject to the
MIT Bag Model boundary condition between two parallel plates in
three spatial dimensions. Afterwards, the Casimir energy has been
calculated for massless fermions in one dimension \,\cite{Jaffe.},
between two parallel plates using various methods in
three-dimensional space
\,\cite{milton.paper.,milton.book.,Massless.,Queiroc.}
and in $d+1$ dimensional space-time \,\cite{Tort2.,Setare.,Bezerra.}. The first
attempt to compute the Casimir energy for the massive case in three spatial
dimensions was done by Mamayev and Trunov who managed to find an integral form for this quantity and explicitly computed its small
and extremely large mass limits \,\cite{Massive.}. The first complete computation of the
Casimir energy for the massive case in this geometry in $3+1$ and $1+1$ dimensions was done by C.
D. Fosco and E. L. Losada in $2008$\,\cite{Functional-approach.}. In
their approach, a coupling of the bilinear form $\bar{\psi} \psi$ to
a series of regularly spaced $\delta$-function potentials with
coupling constant $g$ is introduced, which implements imperfect
bag-like boundary condition. This method can produce a fermionic
propagator which satisfies the MIT Bag Model boundary condition when
$g=2$. However a direct calculation of the Casimir energy for this problem has not been presented so far, and this will be the subject of this paper.
\par
In order to calculate the Casimir energy from first principles one must sum over the
allowed modes. However, the vacuum energies in the presence and
absence of disturbances obtained by these direct sums are infinite.
Therefore one has to adopt regularization and analytic continuation
prescriptions, in order to compute the difference between these two
divergent quantities. When the allowed modes which appear in the
summands are regular, the major approaches used are: the zeta
function analytic continuation technique
\,\cite{milton.paper.,milton.book.,someaspects.,Elizalde2.},
cut-off regularization \,\cite{cut-offs.,Miltao.,CRHagen.,Abel-Plana1.} and
box subtraction scheme along with the Abel-Plana summation
formula\,\cite{Moazzemi.,Abel-Plana1.}. On the other hand, when the
allowed modes which appear in the summands are irregular, the major
approaches used are: the contour integration
method\,\cite{countor.int.}, the Green function
formalism\,\cite{green.func.}, the functional
approach\,\cite{Functional-approach.}, and the Boyer method\,\cite{boyer.,valuyanjadval.}. Unfortunately, most of these
techniques do not be lead to closed forms for the values of the
Casimir energy and one has to employ various numerical methods to
obtain a value for the Casimir energy.
\par
In this paper we calculate the Casimir energy for a massive fermionic field in two dimensional
space-time with the MIT Bag Model boundary condition, by directly summing over the modes. In this problem the fermionic modes are irregular and the divergences that appear are very severe. Upon using the contour integration method, one might encounter some ambiguities, mainly due to the severe nature of divergences inherent to this problem. Here we use a direct approach which does not resort to any analytic continuation techniques and is devoid of any ambiguities. Our method is based on the Boyer method, which we shall explain in detail\,\cite{boyer.}. Moreover, in this procedure one can associate a physical meaning to the Casimir energy: It is equal to the work done in forming the configuration under study from the free vacuum. In section II, we solve the Dirac equation in one spatial dimension using the MIT
Bag Model boundary condition and we find a transcendental equation
for the discrete spectrum. The allowed modes for the massive case
obtained from this equation are not regular. In section III we
compute the Casimir energy by performing a direct sum over all modes
of the field. We finally plot the values obtained for the
Casimir energy as a function of the distance between the points for
various values of the mass. We show that the results for the small
mass limit converges to the results for the massless case. The results obtained in this paper are consistent with those obtained in\,\cite{Functional-approach.}. In Section IV we summarize our results. In the Appendix we present a
rigorous derivation of the MIT Bag Model boundary condition for the
most general case.

\section{The solution of Dirac Equation with the MIT Bag Model boundary condition}
In this section we find the solutions to the Dirac equation with the
MIT Bag boundary condition in one spatial dimension. We consider the
case where the Dirac field is completely free inside the bag:
$(-\frac{a}{2}< x<\frac{a}{2})$. We can write the spinor $\psi(x,t)$
as:
\begin{eqnarray}\label{spinor.}
  \psi(x,t)=e^{-iEt}\left(%
\begin{array}{c}
  f(x) \\
  g(x) \\
\end{array}%
\right),
\end{eqnarray}
where $E$ denotes the energy eigenvalue of the time-independent
solution. We choose the following representation for the
$\gamma$-matrices: $\gamma^{0}=\sigma_{1}$,
$\gamma^{1}=i\sigma_{3}$, and
$\gamma^{5}=\gamma^{0}\gamma^{1}=\sigma_{2}$. Then the Dirac
equation leads to,
\begin{eqnarray}\label{coupled.}
 \Bigg\{\begin{array}{c}
              \hspace{-4cm} f(x)=Ce^{ikx}+De^{-ikx},\\ \\
              g(x)=\frac{m}{E}(Ce^{ikx}+De^{-ikx})+ \frac{i k}{E}(Ce^{ikx}-De^{-ikx}), \\
            \end{array}
\end{eqnarray}
where $k=\sqrt{E^2-m^2}$.
\par
The MIT Bag Model boundary condition has been derived in the
Appendix\,\ref{Appendix.1} and we show that this boundary condition
can completely confine the spinor fields between the boundaries. The
MIT Bag Model boundary condition in two space-time dimension in our
convention becomes,
\begin{eqnarray}\label{B.C.1}
    (1\mp\sigma_3)\psi(x)\bigg|_{x=\pm\frac{a}{2}}=0.
\end{eqnarray}
Applying these conditions we obtain
\begin{eqnarray}\label{B.C.2}
    \bigg\{\begin{array}{c}
              g(\frac{a}{2})=0, \\
              f(\frac{-a}{2})=0. \\
            \end{array}
\end{eqnarray}
\begin{widetext}

  \begin{figure}
  \includegraphics[width=14cm]{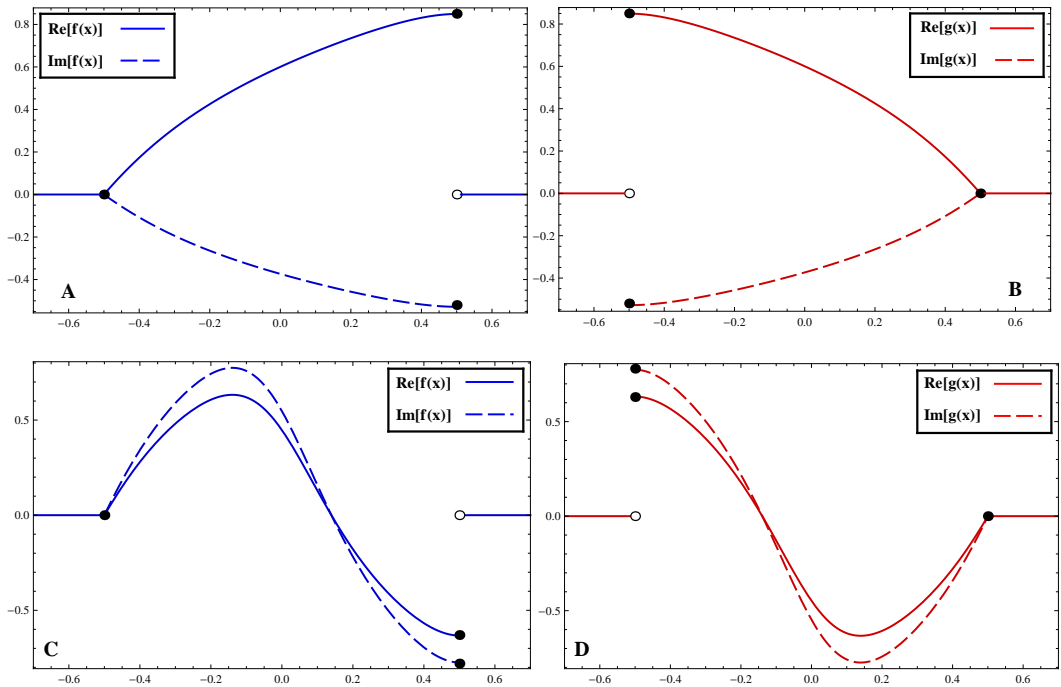}
  \caption{\label{All-Wave-F.} In this figure the real and imaginary parts of the upper and lower components of the lowest two positive energy wavefunctions $\psi(x)$ are plotted as a function of $x$, separately. In parts.\,(A,B) the real and imaginary parts of the upper and lower components of the ground state\,($k_{1}=2.0288$ and $E_1=+\sqrt{k_{1}^2+m^2}$ with $m=1$) are plotted. The parity of this state is positive. In parts.\,(C,D) the real and imaginary parts of the upper and lower components of the first excited state\,($k_{2}=4.9132$ and $E_2=+\sqrt{k_{2}^2+m^2}$ with $m=1$) are plotted. The parity of this state is negative. }
  \end{figure}
\end{widetext}
Extracting $\frac{C}{D}$ from the condition $g(\frac{a}{2})=0$ and inserting it
into $f(\frac{-a}{2})=0$, we obtain an expression which determines
the quantized modes:
\begin{eqnarray}\label{B.C.3}
    ka\cot(ka)=-ma.
\end{eqnarray}
The above expression is identical to the result obtained by Mamayev
and Trunov for the component of the momentum perpendicular to the
plates for a massive Dirac field between two parallel plates in
three spatial dimensions\,\cite{Massive.}. In
Ref.\,\,\cite{Massive.} this condition has been obtained by using
the MIT boundary condition and the property of the Dirac spinors
that each component also satisfies the Klein-Gordon equation.
However all of the allowed modes along with this equation for the
two parallel plate geometry in 3+1 dimension can also be easily
obtained directly, along the same lines as the derivation presented
in this section. The solutions of the above transcendental
equation\,(Eq.\,(\ref{B.C.3})), denoted by $k_{s}$, are not regular
for the massive Dirac fields. Only for the specific case of a
massless Dirac field, the modes are regular ($k_s a=
\frac{(2s-1)\pi}{2}$, $s=\{1,2,3, \cdots \}$)
\,\cite{milton.paper.,milton.book.}. Since changing the sign of the root $k_{s}$ does not lead to a linearly independent solution, we consider only the positive roots.
\par
As expected the MIT bag model boundary condition has transformed the spectrum of the free Dirac field, which consisted of two continua starting at $E=\pm m$, into two sets of discrete states with energies $E=\pm \sqrt{k_{s}^2+m^2}$, where $s=\{1,2,3,\cdots\}$. This problem has parity symmetry and in this representation the parity operation is given by $ P\psi(x,t)=\sigma_1 \psi(-x,t)$. The energy eigenstates automatically turn out to be parity eigenstates. The parities of the lowest lying states with energies $E=\pm \sqrt{k_{1}^2+m^2}$ are $\pm 1$ respectively, and the parities of the exited states alternate as the absolute value of the energy increases. The real and imaginary parts of the upper and lower components of the lowest two positive energy wavefunctions $\psi(x)$ are plotted in Fig.\,(\ref{All-Wave-F.}). Note that the values of the wavefunctions on the boundaries are non-zero and just outside the boundaries are exactly zero. This is another manifestation of the imposition of the MIT bag model boundary condition, as explained fully in the Appendix\,\ref{Appendix.1}.
\par
In order to calculate the Casimir energy for the massive case we need to find a relationship
between the root number $s$ and the wave number $k_{s}$. For this
purpose, Eq.\,(\ref{B.C.3}) can be written as:
\begin{eqnarray}\label{relationship1.}
  s=\frac{1}{\pi}\Big[X_{s}+\tan^{-1}(\frac{X_{s}}{M})\Big],
\end{eqnarray}
where $s$ is the root number, $X_s=k_s a$ and $M=ma$. Note that the values of the
root indices $s$ and the corresponding wave-numbers $k_{s}$ can be
analytically continued to any real value.
\section{The Casimir Energy}
In this section we calculate the Casimir energy for a massive Dirac
field between two parallel plates\,(two points) in 1+1 space-time
dimensions. In order to obtain the Casimir energy, we should subtract the zero point energy in the absence from
the presence of the boundary conditions. Therefore, the vacuum energies for both cases should be calculated.
Since the solutions of the Hamiltonian in the absence and presence of the boundaries are complete \cite{mackenzie1,dr}, the Fermi field operator can be expanded in terms of either of these modes, as follows
\begin{eqnarray}\label{wave.f.}
\dd\Xi(x)&=&\int_{-\infty}^{\infty}\frac{dk}{2\pi}[b_k u_k(x)
+d^{\dag}_{k} \upsilon_{k} (x)]
\nonumber\\ \dd&=&\sum_{s=1}^{\infty}[a_{s}\mu_{k_s}(x)+c^{\dag}_{s} \nu_{k_s} (x)]
\end{eqnarray}
where we have denoted the positive-energy and negative-energy modes in the free case by $u_{k}(x)$ and $\upsilon_{k}(x)$, and in the case with the boundary condition by $\mu_{k_s}(x)$ and $\nu_{k_s} (x)$, respectively.  By substituting the expressions for the field operator $\Xi(z)$ given in Eq.\,(\ref{wave.f.}) into the general definition of the Hamiltonian operator and using the usual anticommutation relations, and evaluating the two zero point energies, we obtain the following expression for the Casimir energy,
\begin{eqnarray}\label{zero.p.}
\dd &E&_{\hspace{-0.2cm}\footnotesize\mbox{Cas.}}=\langle\Omega\mid H\mid\Omega\rangle-\langle0\mid H^{\footnotesize\mbox{free}}\mid0\rangle\nonumber\\
 \dd &=&\int_{-\infty}^{+\infty}dx\sum_{s=1}^{\infty}(-\sqrt{k_s^2 +m^2})\nu^{\dag}_{k_s} (x) \nu_{k_s} (x) \nonumber\\
\dd &-&\int_{-\infty}^{+\infty}dx\int_{-\infty}^{+\infty}\frac{adk}{2\pi}(-\sqrt{k^2 +m^2})\upsilon^{\dag}_{k} (x)\upsilon_{k} (x),
\end{eqnarray}
where $\mid0\rangle$ and $\mid\Omega\rangle$ denote the vacuum states in the absence and the presence of the boundary condition, respectively. We have just shown that we can obtain the vacuum energy, and therefore the Casimir energy, by simply summing over the negative energy modes without the factor of $1/2$. This is equivalent to the usual definition where one sums over both positive and negative energy modes, since our problem possesses particle conjugation symmetry along with C, P and T symmetries, separately. Integrating over $x$ the Casimir energy becomes,
\begin{eqnarray}\label{newenergy.}
    E_{\footnotesize\mbox{Cas.}}=-\sum_{s=1}^{\infty} \sqrt{k_{s}^2+m^2}+\int_{-\infty}^{+\infty}\frac{adk}{2\pi}\sqrt{k^2 +m^2}.
\end{eqnarray}
As mentioned earlier both of these vacuum energies are infinite. However the difference, which is the Casimir energy, is expected to be finite. In many of the techniques for calculating the Casimir energy one starts with only the expression for the vacuum energy for the problem at hand (the first expression in Eq.\,(\ref{newenergy.}) in our case) and removes the infinities that appear during the calculation by using various methods such as analytic continuation or simply by hand. this should precisely amount to subtracting the free vacuum energy which was left out from the beginning (the second expression in Eq.\,(\ref{newenergy.}) in our case).
\par
\begin{figure}[th]
 \hspace{-0.7cm} \includegraphics[width=8.8cm]{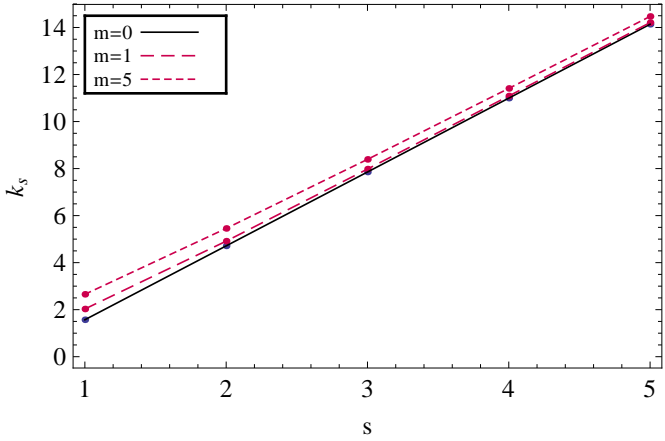}
\caption{\label{massive-massless2.} The plot of the allowed values
of the wave-number $k_{s}$, obtained from the roots of
Eq.\,(\ref{B.C.3}), as a function of the root number $s$. We have
displayed the results for three values of the mass $m=\{0,1,5\}$ with $a=1$.
Note that for $m=0$ we obtain a straight line, \textit{i.e.} the roots are
equally spaced. However, as is apparent from the figure, this no
longer true for $m \neq 0$. }
\end{figure}

The dependence of the fermionic quantized momenta on the mass of the
field is one of the distinguishing features of the Fermi field as
compared to the bosonic case. In Fig.\,(\ref{massive-massless2.}) we show the wave vectors for two
massive and a massless fermionic field.
However as mentioned earlier, for $m=0$ the wave vectors are evenly spaced and are given by $k_s=\frac{(2s-1)\pi}{2a}$. Therefore the Casimir energy for a
massless fermionic field can be easily obtained from the first term in Eq.\,(\ref{newenergy.}) using the zeta function and its analytic
continuation as follows\,\cite{milton.book.,Jaffe.},
\begin{eqnarray}\label{zetafunction.}
   E_{\mbox{\footnotesize {Cas.}}}^{(0)}\Big(M=0\Big)
   &=&\frac{-1}{a}\Bigg[\sum\limits_{s=1}^{\infty}(s-\frac{1}{2})\pi\Bigg]_{\mbox{\footnotesize {Analytic Part}}}\nonumber\\
     &=&\frac{-\pi}{a}\zeta(-1,\frac{1}{2})=\frac{-\pi}{24a}.
\end{eqnarray}
\par
\begin{figure}[th]
 \hspace{-0.1cm} \includegraphics[width=7cm]{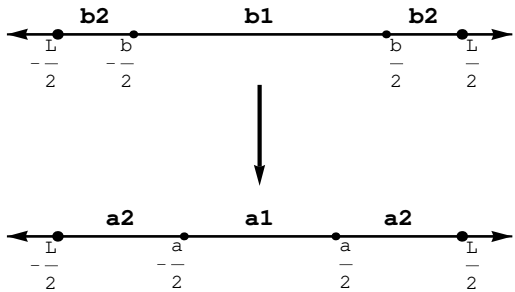}
\caption{\label{Box.fig1.} The geometry of the two different
configurations whose energies are to be compared. The labels $a1$,
$b1$, etc. denote the appropriate sections in each configuration
separated by points. The upper configuration is denoted by `B', and
the lower one by `A'.}
\end{figure}
\begin{figure}[th]
 \hspace{-0.7cm} \includegraphics[width=8.8cm]{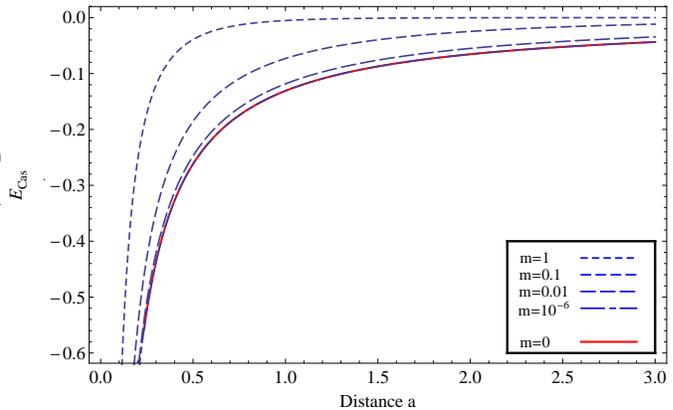}
\caption{\label{fig.as.a} The values of the Casimir energy for the
massive and massless Dirac fields with the MIT Bag Model boundary
condition in one spatial dimension are plotted as a function of the
distance between the points\,($a$). In this figure we have shown a
sequence of plots for $m=\{1,0.1,0.01,10^{-6},0\}$. It is apparent
that the sequence of the plots for the massive cases converges
rapidly to the massless case as $m$ decreases.}
\end{figure}

As shown in Fig.\,(\ref{massive-massless2.}) the wave numbers
obtained from Eq.\,(\ref{B.C.3}) for a massive Dirac field are
irregular. In order to calculate the Casimir energy we use the Boyer method\,\cite{boyer.}, instead of using Eq.\,(\ref{newenergy.}) directly, since the latter is more prone to ambiguities. These two methods for calculating of the Casimir energy are equivalent. Now we discuss the Boyer method in detail. In this method we consider two similar configurations: two points
with distance $a$ and two points with distance $b$. Then we place
each of these systems in a box with size $L>a,b$ as
Fig.\,(\ref{Box.fig1.}). Finally we define the Casimir energy as the
difference between the vacuum energies of these two similar
configurations as follows,
\begin{eqnarray}\label{Casimir.def.}
   E_{\mbox{\footnotesize {Cas.}}}=\lim_{b/a \rightarrow \infty}\bigg[\lim_{L/b \rightarrow \infty}\big[E_{A}^{(0)}-E_{B}^{(0)}\big]\bigg],
\end{eqnarray}
where $E_{A}^{(0)}=E_{a1}^{(0)}+2E_{a2}^{(0)}$ and
$E_{B}^{(0)}=E_{b1}^{(0)}+2E_{b2}^{(0)}$ are the vacuum energies of
configurations `A' and `B' as shown in Fig.\,(\ref{Box.fig1.}). Note that upon taking the limits indicated in Eq.\,(\ref{Casimir.def.}), one recovers the original definition of the Casimir energy given in Eq.\,(\ref{newenergy.}). From
this definition one can easily conclude that the Casimir energy is
equal to the work done on the configuration `B' to deform it to
configuration `A'. For each of the six regions shown in the
Fig.\,(\ref{Box.fig1.}), for example the region $a1$, we calculate
the vacuum energy as follows,
\begin{eqnarray}\label{vcum.ergy.conv.fact.}
   E^{\mbox {\footnotesize $(0)$}}_{a1}(M)=\lim_{\lambda\rightarrow 0}\bigg[-\frac{1}{a}\sum_{s=1}^{\infty} (X_{s}^2+M^2)^{1/2}g(\lambda
   \omega_{s})\bigg],
\end{eqnarray}
where we have introduced a convergence factor $g(\lambda
\omega_{s})=e^{-\lambda \sqrt{X_{s}^{2}+M^{2}}}$ with
$\lambda\rightarrow 0$. Now we start the calculation for the region
$a1$, and the calculations for the other regions can be done
analogously. Since the wave-vectors are not regular with respect to
the root indices $s$, in order to find an analytical form for the
divergence in Eq.\,(\ref{vcum.ergy.conv.fact.}), we employ the
simplest form of the Euler-Maclaurin Summation Formula\,(EMSF) to
obtain
\begin{eqnarray}\label{vcum.ergy.EMSF1.}
   \dd E^{\mbox {\footnotesize $(0)$}}_{a1}(M)&=&-\frac{1}{a}\lim_{\lambda\rightarrow 0}\Big[\int_{s=1}^{\infty}ds(X_{s}^{2}+M^2)^{1/2}g(\lambda \omega_{s})\nonumber\\&&\hspace{-0.7cm}+\frac{1}{2}(X_{1}^{2}+M^2)^{1/2}g(\lambda \omega_{1})\\&&\hspace{-0.7cm}+\int_{s=1}^{\infty}ds\Big(s-[s]-\frac{1}{2}\Big)\frac{d}{ds}[(X_{s}^{2}+M^2)^{1/2}g(\lambda \omega_{s})]\Big],\nonumber
\end{eqnarray}
where $[s]$ is the Floor function. There is a one-to-one
correspondence between the wave vector $k_{s}$ and the root number
$s$, as is manifest in Eq.\,(\ref{relationship1.}). Therefore we can
change the variable of integration from $s$ to $X$. Then we obtain,
\begin{eqnarray}\label{vcum.ergy.EMSF.after.cv.}
   \dd E^{\mbox {\footnotesize $(0)$}}_{a1}(M)&=&-\frac{1}{a}\lim_{\lambda\rightarrow 0}
   \Big[\int_{X=X_{1}}^{\infty}dX\frac{ds}{dX}(X^{2}+M^2)^{1/2}g(\lambda \omega)\nonumber\\&&\hspace{-1.3cm}+
   \frac{1}{2}(X_{1}^{2}+M^2)^{1/2}g(\lambda \omega_{1})\\&&\hspace{-1.3cm}+\int_{X=X_{1}}^{\infty}dX
   \Big(s-[s]-\frac{1}{2}\Big)\frac{d}{dX}[(X^{2}+M^2)^{1/2}g(\lambda \omega)]\Big].\nonumber
\end{eqnarray}
Now, by adding and subtracting appropriate terms, we can extend
the lower limits of all integrals in
Eq.\,(\ref{vcum.ergy.EMSF.after.cv.}) to zero. We thus have
\begin{eqnarray}\label{vcum.ergy.EMSF.ext0.}
   \dd E^{\mbox {\footnotesize $(0)$}}_{a1}(M)&=&-\frac{1}{a}\lim_{\lambda\rightarrow 0}\Big[\int_{X=0}^{\infty}dX\frac{ds}{dX}(X^{2}+M^2)^{1/2}g(\lambda \omega)
   \nonumber\\&&\hspace{-1.3cm}-\int_{X=0}^{X_{1}}dX\frac{ds}{dX}(X^{2}+M^2)^{1/2}g(\lambda \omega)\nonumber\\&&\hspace{-1.3cm}+\frac{1}{2}(X_{1}^{2}+M^2)^{1/2}g(\lambda \omega_{1})
   \\&&\hspace{-1.3cm}+\int_{X=0}^{\infty}dX\Big(s-[s]-\frac{1}{2}\Big)\frac{d}{dX}[(X^{2}+M^2)^{1/2}g(\lambda \omega)]\nonumber\\&&\hspace{-1.3cm}
   -\int_{X=0}^{X_{1}}dX\Big(s-[s]-\frac{1}{2}\Big)\frac{d}{dX}[(X^{2}+M^2)^{1/2}g(\lambda \omega)]\Big].\nonumber
\end{eqnarray}
The last term in Eq.\,(\ref{vcum.ergy.EMSF.ext0.}) can be simplified
by noting that the Floor function $[s]=0$ in the indicated domain,
and integration by parts yields,
\begin{eqnarray}\label{Int.by.part.}
   &&\hspace{-1.5cm}\int_{X=0}^{X_{1}}dX\Big(s-[s]-\frac{1}{2}\Big)\frac{d}{dX}[(X^{2}+M^2)^{1/2}g(\lambda \omega)]\nonumber\\&=&
   \int_{X=0}^{X_{1}}dX\Big(s-\frac{1}{2}\Big)\frac{d}{dX}[(X^{2}+M^2)^{1/2}g(\lambda \omega)]\nonumber\\&=&\frac{1}{2}(X_{1}^{2}+M^2)^{1/2}g(\lambda \omega_{1})+\frac{1}{2}M e^{-\lambda M}\nonumber\\ &-&\int_{X=0}^{X_{1}}dX\frac{ds}{dX}(X^{2}+M^2)^{1/2}
   g(\lambda \omega).\nonumber\\
\end{eqnarray}
Using Eqs.\,(\ref{vcum.ergy.EMSF.ext0.},\ref{Int.by.part.}) we
obtain
\begin{eqnarray}\label{vcum.ergy.EMSFfinal.}
   \dd &&\hspace{-0.5cm}E^{\mbox {\footnotesize $(0)$}}_{a1}(M)=\\
   \hspace{-2.5cm}&-&\frac{1}{a}\lim_{\lambda\rightarrow 0}
   \Bigg[\int_{X=0}^{\infty}dX\frac{ds}{dX}(X^{2}+M^2)^{1/2}g(\lambda \omega)+\frac{1}{2}M e^{-\lambda
   M}\nonumber\\&+&\int_{X=0}^{\infty}dX\Big(s-[s]-\frac{1}{2}\Big)\frac{d}{dX}
   [(X^{2}+M^2)^{1/2}g(\lambda \omega)]\Bigg].\nonumber
\end{eqnarray}
Note that only the first term on the right hand side of
Eq.\,(\ref{vcum.ergy.EMSFfinal.}) is divergent. Upon substituting
the expression displayed in Eq.\,(\ref{vcum.ergy.EMSFfinal.}) into
the definition of the Casimir energy given in
Eq.\,(\ref{Casimir.def.}), the constant terms automatically cancel
each other in the limit $\lambda\rightarrow 0$ and by choosing
appropriate cutoffs on the upper limits of the integrals for each
region, the divergent integrals cancel each other, all due to the
box subtraction scheme. Therefore only the convergent integral terms
remain. The contributions of integrals in regions $a2$, $b1$ and $b2$ go
to the zero as $L/b\rightarrow\infty$ and $b/a\rightarrow\infty$.
Therefore our final expression for the Casimir energy is,
\begin{eqnarray}\label{cas.ergy.EMSFfinal.}
   \dd E_{\mbox {\footnotesize Cas.}}=\lim_{\lambda\rightarrow 0}\Bigg[-\frac{1}{a}\int_{0}^{\infty}dX\Big(s-[s]-\frac{1}{2}\Big)
   \nonumber\\ \times\frac{d}{dX}[(X^{2}+M^2)^{1/2}g(\lambda \omega)]\Bigg],
\end{eqnarray}
where $s$ is obtained from Eq.\,(\ref{relationship1.}). It seems
that this expression does not have a closed form solution and it should be
solved numerically. As is apparent from
Eq.\,(\ref{cas.ergy.EMSFfinal.}), the integrand has an infinite
number of discontinuities due to the presence of the Floor function.
First the precise positions of the jumps in the integrand have to be
determined. These jumps precisely correspond to the roots of
Eq.\,(\ref{B.C.3}), giving the values of the wave-numbers. The
integrations are done separately for all parts and then all of the
results are summed. The integration is over the continuous version
of the wave number, which extends to infinity. In order to
accomplish this numerically, we compute this integral up to a cutoff
$\Lambda$ which should eventually go to infinity. Meanwhile, we also
have to take the limit $\lambda\rightarrow 0$ as indicated in
Eq.\,(\ref{cas.ergy.EMSFfinal.}). We have determined that an
optimization occurs precisely when $\lambda=1/\Lambda$.
In Fig.\,(\ref{fig.as.a}) the values of the Casimir energy have been
plotted as a function of the distance $a$ for various values of $m$.
This plot shows that there is a good consistency between the results
of the massless case and massive ones when $m\rightarrow 0$. This
figure also shows the rapid decrease in the value of the Casimir
energy as a function of $ma$. We should mention that our results are
in an excellent agreement with the previously reported result which
was obtained indirectly through the analysis of the bilinear
forms\,\cite{Functional-approach.}.

\section{Conclusion}
In this paper we have computed the Casimir energy for a massive
Dirac field with the MIT Bag Model boundary condition in one spatial
dimension. We have used the direct mode summation method in order to
compute the Casimir energy for this field. For the massless case the
modes are regular and we use the zeta function analytic continuation
technique. However, for the massive case the modes are irregular and
we use the Boyer's subtraction scheme. Its worth mentioning that in
this technique all of the infinities are canceled automatically and
there is no need to use any analytic continuation. We have shown
that the massless limit of the massive case precisely corresponds to
the massless case. Our result for the values of the Casimir energy
has been obtained numerically, similar to the previously reported
results\,\cite{Functional-approach.}.

\section*{Acknowledgement} We would like to thank the Research Office
of the Shahid Beheshti University for financial support.

\appendix
\section{A Derivation for the MIT Bag Model Boundary condition} \label{Appendix.1}
In this appendix we present a rigorous derivation of the MIT Bag
Model boundary condition for the Dirac field inside an arbitrary
closed surface $S$. This boundary condition ensures the complete
confinement of the eigenstates of the Dirac Hamiltonian inside an
enclosed area. We show that this boundary condition can be obtained
by coupling the Dirac field to a scalar potential $V$ and taking the
limit as $V\rightarrow \infty$. However we first present the reasons
why we cannot confine fermions inside an enclosed area by the time
component of a four-vector potential $V_0$.
\par
Solving the Dirac equation we obtain $p^2=(E-V_0+m)(E-V_0-m)$ which
is positive for $|E-V_0|>m$. Therefore we have oscillatory solutions
out of the barrier, i.e. we have currents of particles and
antiparticles. This contradicts the assumption of complete
confinement of the fermionic field. This is the well-known Klein's
paradox \,\cite{Klein.}. Next we use the scalar potential $V(x)$
which, as we shall see, does not have this problem. The Dirac
equation with the scalar potential is:
\begin{eqnarray}\label{Dirac-equation.}
    \big[i\gamma^\mu \partial_{\mu}-(m+V(\vec{x}))\big] \psi(\vec{x},t)=0.
\end{eqnarray}
Decomposing the spatial components of $\gamma^\mu \partial_{\mu}$ at
the surface into tangential $(t)$ and normal $(n)$ parts, we have:
\begin{eqnarray}\label{gama.}
    \gamma^\mu \partial_{\mu}=\gamma^{0} \partial_{0}+\gamma^{t}
    \partial_{t}+\gamma^{n} \partial_{n}.
\end{eqnarray}
\par
We choose the potential $V(\vec{x})$ to be infinite outside of the
enclosed area and to vanishes inside. First we take the integral of
the Dirac equation with the scalar potential
(Eq.\,(\ref{Dirac-equation.})) from $\vec{a}-\epsilon \hat{n}$ to
$\vec{a}$ (a small interval inside the barrier) where $\vec{a}$
specifies a random point on the surface and $\hat{n}$ is the normal
unit vector to the surface at point $\vec{a}$
\begin{eqnarray}\label{integral1.}
    \int_{a-\epsilon}^{a}\hat{n}dn(i\gamma^\mu
    \partial_{\mu}-m)\psi(\vec{x},t)=0.
\end{eqnarray}
When $\epsilon\rightarrow{0}$, all of the terms vanish except the
one containing the term $\gamma^n \partial_n$. Then:
\begin{eqnarray}\label{shart1.}
    i(\hat{n}\cdot\vec{\gamma})(\psi(\vec{a})-\psi(\vec{a}-\epsilon\hat{n}))=0 \Rightarrow
    \psi(\vec{a})=\psi(\vec{a}-\epsilon \hat{n}).
\end{eqnarray}
Second, we change the integration domain. This time we integrate
from a point just inside the volume ($\vec{a}-\epsilon \hat{n}$) to
a point outside of the volume ($\vec{a}+\epsilon \hat{n}$).
\begin{eqnarray}\label{integral2.}
    \int_{a-\epsilon}^{a+\epsilon}\hat{n}dn(i\gamma^\mu
    \partial_{\mu}-(m+V(\vec{x})))\psi(\vec{x},t)=0.
\end{eqnarray}
This time we have to deal with the infinite potential outside the
bag and therefore we cannot neglect the term $\epsilon V$. Now using
the fact that the Dirac equation demands
$\psi(\vec{a}+\epsilon\hat{n})=0$ and Eq.\,(\ref{shart1.}) one
obtains:
\begin{eqnarray}\label{shart2.}
    -i(\hat{n}\cdot\vec{\gamma})\psi(\vec{a})=\epsilon V\psi(\vec{a}).
\end{eqnarray}
Multiplying Eq.\,(\ref{shart2.}) from left by
$(\hat{n}\cdot\vec{\gamma})$ yields $\epsilon V=\pm1$. Since
$\epsilon>0$ and $V=+\infty$ we conclude $\epsilon V=1$.
\par
It is interesting to note that using the time component of a
four-vector potential for the confinement purpose we obtain an
inconsistent result: $\epsilon V=\pm i$.
\par
Inserting $\epsilon V=1$ in Eq.\,(\ref{shart2.}) we obtain the MIT
Bag Model boundary condition:
\begin{eqnarray}\label{bag-boundary.}
    (1+i(\hat{n}\cdot\vec{\gamma}))\psi(\vec{x})\bigg|_{\mbox {\footnotesize Boundary}}=0.
\end{eqnarray}



\begin{thebibliography}{9}
\bibitem{h.b.g.}
     H. B. G. Casimir, Proc. Kon. Nederl. Akad. Wet. \textbf{51} (1948) 793.
\bibitem{burdag.book.}
     M. Bordag, G.L. Klimchitskaya, U. Mohideen, and V.M. Mostepanenko, \emph{Advances in the Casimir Effect}, Oxford University Press Inc. New York (2009).
\bibitem{milton.book.}
     K. A. Milton, \emph{The Casimir Effect: Physical Manifestations of Zero-Point Energy}, (World Scientific Publishing Co. Pte. Ltd. 2001).
\bibitem{milton.paper.}
     K. A. Milton, \emph{The Casimir Effect: Physical Manifestations of Zero Point Energy},
     Invited Lectures at 17th Symposium on Theoretical Physics, Seoul National University, Korea, June 29-July 1,
     (1998); [{\tt arXiv:hep-th/9901011v1}].
\bibitem{Mostepanenko.}
     M. Bordag, U. Mohideen, V.M. Mostepanenko, Phys. Rep. \textbf{353} (2001) 1; [{\tt arXiv:quant-ph/0106045v1}].
\bibitem{chiral.casimir.}
    Hee-Jung Lee, Dong-Pil Min, Byung-Yoon Park, Mannque Rho and Vicente Vento, Nucl. Phys. A \textbf{657} (1999) 75.
\bibitem{linas.}
    Linas Vepstas, A. D. Jackson and A. S. Goldhaber, Nucl. Phys. B \textbf{140} (1984) 280.
\bibitem{bag1.}
     I. O. Cherednikov, Int. J. Mod. Phys. A \textbf{17}, 874 (2002).
\bibitem{LBrink.}
     L. Brink and H. B. Nielsen, Phys. Lett. B \textbf{45}, 332 (1973).
\bibitem{condensed.}
     F. De Martini, M. Marrocco and P. Mataloni, Phys. Rev. A \textbf{43}, 2480 (1991).
\bibitem{Krech.}
     M. Krech and S. Dietrich, Phys. Rev. Lett. \textbf{66}, 345 (1991).
\bibitem{nano.application.}
     F. Chen, U. Mohideen, G. Klimchitskaya and V. Mostepanenko, Phys. Rev. Lett. \textbf{88} (2002) 101801.
\bibitem{Rodrigues.}
     R. Rodrigues, P. Maia Neto, A. Lambrecht and S. Reynaud, EPL \textbf{76}, 822 (2006).
\bibitem{Lambrecht.}
     A. Lambrecht, I. Pirozhenko, L. Duraffourg and Ph. Andreucci, EPL \textbf{77}, 44006 (2007).
\bibitem{5.}
     G.W. Semenoff, Phys. Rev. Lett. \textbf{53}, (1984) 2449.
\bibitem{Vincenzo.}
     D.P. Di Vincenzo and E.J. Mele, Phys. Rev. B \textbf{29} (1984) 1685.
\bibitem{Gonzalez.}
     J. Gonz\'{a}lez, F. Guinea, and M. A. H. Vozmediano, Phys. Rev. B \textbf{63} (2001) 134421.
\bibitem{Lee.}
     H.-W. Lee and D.S. Novikov, Phys. Rev. B \textbf{68} (2003) 155402.
\bibitem{Sharapov.}
     S.G. Sharapov, V.P. Gusynin, and H. Beck, Phys. Rev. B \textbf{69} (2004) 075104.
\bibitem{Castro.}
     A.H. Castro Neto, F. Guinea, N.M.R. Peres, K.S. Novoselov, and A.K. Geim, Rev. Mod. Phys. \textbf{81} (2009) 109.
\bibitem{graphene.ref.}
    S. Bellucci and A. A. Saharian, Phys. Rev. D\textbf{80} (2009)  105003; [{\tt arXiv:hep-th/0907.4942v1}]
\bibitem{EM.}
     J. Ambj{\o}rn, and S. Wolfram, Ann. Phys. (N. Y.) \textbf{147} (1983) 1.
\bibitem{Valuyan.}
     M. A. Valuyan, R. Moazzemi, and S. S. Gousheh, J. Phys. B: At. Mol. Opt. Phys. \textbf{41} (2008) 145502; [{\tt arXiv:hep-th/0806.1628}].
\bibitem{Hacgan.}
     S. Hacgan, R. J\'{a}uregui, and C. Villarreal, Phys. Rev. A \textbf{47} (1993) 4204.
\bibitem{Jordan.}
     G. Jordan Maclay, Phys. Rev. A \textbf{61} (2000) 052110-1.
\bibitem{valuyanjadval.}
     M. A. Valuyan, and S. S. Gousheh, Int. J. Mod. Phys. A \textbf{25} (2010) 1165.
\bibitem{Moazzemi.}
    R. Moazzemi, M. Namdar, and S.S. Gousheh, JHEP \textbf{09} (2007) 029; [{\tt arXiv:hep-th/0708.4127v1}].
\bibitem{Gousheh.}
    R. Moazzemi, and S. S. Gousheh, Phys. lett. B \textbf{658} (2008) 255; [{\tt arXiv:hep-th/0708.3428v2}].
\bibitem{valuyan2.}
    S. S. Gousheh, R. Moazzemi, and M. A. Valuyan, Phys. Lett. B \textbf{681} (2009) 477; [{\tt arXiv:hep-th/0911.3707}].
\bibitem{Mhammadi.}
    R. Moazzemi, Abdollah Mohammadi, and S.S. Gousheh, Eur. Phys. J. C \textbf{56} (2008) 585; [{\tt arXiv:hep-th/0806.4862}].
\bibitem{mixed.}
     A. C. Aguiar Pinto, T. M. Britto, R. Bunchaft, F. Pascoal, and F.S.S. da Rosa, Brazilian Journal of Physics \textbf{33} (2003) 60.
\bibitem{Robins.}
     A. Romeo, and A. A. Saharian, J. Phys. A: Math. Gen. \textbf{35} (2002) 1297; [{\tt arXiv:hep-th/0007242v2}].
\bibitem{Albuquerque.}
     L.C. de Albuquerque, and R. M. Cavalcanti, J. Phys. A: Math. Gen. \textbf{37} (2004) 7039; [{\tt arXiv:hep-th/0311052v2}].
\bibitem{someaspects.}
     C. Farina, Brazilian Journal of Physics \textbf{36} (2006) 1137.
\bibitem{Jaffe.}
     P. Sundberg, and R. L. Jaffe, Ann. Phys. \textbf{309} (2004) 449; [{\tt arXiv:hep-th/0308010v1}].
\bibitem{Bogolioubov.}
     P. N. Bogolioubov, Ann. Inst. Henri Poincare \textbf{8} (1967) 163.
\bibitem{Chodos1.}
     A. Chodos, R. L. Jaffe, K. Johnson, C. B. Thorn, and V. F. Weisskopf, Phys. Rev. D \textbf{9} (1974) 3471.
\bibitem{Chodos2.}
      A. Chodos, R. L. Jaffe, K. Johnson, and C. B. Thorn, Phys. Rev. D \textbf{10} (1974) 2599.
\bibitem{Elizalde.}
     E. Elizalde, M. Bordag, and K. Kirsten, J. Phys. A: Math. Gen. \textbf{31} (1998) 1743; [{\tt arXiv:hep-th/9707083v1}].
\bibitem{hofmann.}
     R. Hofmann, M. Schumann and R.D. Viollier, Eur. Phys. J. C \textbf{11} (1999) 153.
\bibitem{maghale.}
     A. Seyedzahedi, R. Saghian, and S. S. Gousheh, Phys. Rev. A.\textbf{82} (2010) 032517.
\bibitem{Massless.}
     K. Johnson, Acta. Pol. B \textbf{6} (1975) 865.
\bibitem{Queiroc.}
    H. Queiroc, J. C. da Silva, and F. C. Khanna, J.M.C. Malbouisson, M. Revzen and A. E. Santana, Ann. Phys. \textbf{317} (2005) 220.
\bibitem{Tort2.}
     E. Elizalde, F. C. Santos, and A. C. Tort, Int. J. Mod. Phys. A \textbf{18} (2003) 1761; [{\tt arXiv:hep-th/0206114v1}].
\bibitem{Setare.}
     A. A. Saharian and M. R. Setare, Int. J. Mod. Phys. A \textbf{19} (2004) 4301.
\bibitem{Bezerra.}
     A. A. Saharian and E. R. Bezerra de Mello, Int. J. Mod. Phys. A \textbf{20} (2005) 2380.
\bibitem{Massive.}
     S. G. Mamayev and N. N. Trunov, Sov. Phys. J \textbf{23} (1980) 551.
\bibitem{Functional-approach.}
     C. D. Fosco, and E. L. Losada, Phys. Rev. D \textbf{78} (2008) 025017; [{\tt arXiv:hep-th/0805.2922v1}].
\bibitem{Elizalde2.}
     E. Elizalde, S. D. Odindsov, A. Romeo, A. A. Bitsenko and S. Zerbini, \emph{Zeta Regularization Techniques with Appliccations}, (World Scientific, Singapore, 1994).
\bibitem{cut-offs.}
     N. F. Svaiter, and B. F. Svaiter, J. Phys. A: Math. Gen. \textbf{25} (1992) 979.
\bibitem{Miltao.}
     M. S. R. Milt$\tilde{a}$o, Phys. Rev. D \textbf{78} (2008) 065023.
\bibitem{CRHagen.}
     C.R. Hagen, Eur. Phys. J. C \textbf{19} (2001) 677.
\bibitem{Abel-Plana1.}
     A. A. Saharian, \emph{The generalized Abel-Plana formula: applications to Bessel functions and casimir effect}, [{\tt arXiv:hep-th/0002239v1}].
\bibitem{countor.int.}
     V. V. Nesterenko and I. G. Pirozhenko, Phys. Rev. D \textbf{57} (1998) 2.
\bibitem{green.func.}
     K. A. Milton, L. L. Deraad, and J. Schwinger, Ann. Phys. (N.Y.) \textbf{115}, 388 (1978).
\bibitem{boyer.}
     T. H. Boyer, Phys. Rev.\textbf{174} (1968) 1764.
\bibitem{Klein.}
     J. J. Sakurai, \emph{Advanced Quantum Mechanics}, Addison Wesley; Rev sub edition (1993).
\bibitem{mackenzie1}
R. MacKenzie and F. Wilczek, Phys. Rev. D \textbf{30}, 2194 (1984).
\bibitem{dr}
S. S. Gousheh and R. L\'{o}pez-Mobilia, Nucl. Phys. B \textbf{428},
189 (1994).

\end{thebibliography}
 \end{document}